\documentclass[aps,prl,10pt,twocolumn,showpacs,preprintnumbers,amsmath,amssymb,superscriptaddress]{revtex4-1}
\usepackage[english]{babel}
\usepackage[dvips]{graphicx}
\usepackage{dcolumn}
\usepackage{bm}

\usepackage{color}

\begin{document}

\author{M. E. Dieckmann}
\affiliation{Department of Science and Technology (ITN), Link\"opings University, Campus Norrk\"oping, SE-60174 Norrk\"oping, Sweden}

\author{G. Sarri}
\affiliation{Centre for Plasma Physics (CPP), Queen's University Belfast, BT7 1NN, UK}

\author{D. Folini}
\affiliation{\'Ecole Normale Sup\'erieure, Lyon, CRAL, UMR CNRS 5574, Universit\'e de Lyon, France}

\author{R. Walder}
\affiliation{\'Ecole Normale Sup\'erieure, Lyon, CRAL, UMR CNRS 5574, Universit\'e de Lyon, France}

\author{M. Borghesi}
\affiliation{Centre for Plasma Physics (CPP), Queen's University Belfast, BT7 1NN, UK}

\date{\today}
\email{mark.e.dieckmann@liu.se}
\title{Cocoon formation by a mildly relativistic pair jet in unmagnetized collisionless electron-proton plasma}
\begin{abstract}
By modelling the expansion of a cloud of electrons and positrons with the temperature 400 keV that propagates at the mean speed 0.9c ($c:$ speed of light) through an initially unmagnetized electron-proton plasma with a particle-in-cell (PIC) simulation, we find a mechanism that collimates the pair cloud into a jet. A filamentation (beam-Weibel) instability develops. Its magnetic field collimates the positrons and drives an electrostatic shock into the electron-proton plasma. The magnetic field acts as a discontinuity that separates the protons of the shocked ambient plasma, known as the outer cocoon, from the jet's interior region. The outer cocoon expands at the speed 0.15c along the jet axis and at 0.03c perpendicularly to it. The filamentation instability converts the jet's directed flow energy into magnetic energy in the inner cocoon. The magnetic discontinuity cannot separate the ambient electrons from the jet electrons. Both species rapidly mix and become indistinguishable. The spatial distribution of the positive charge carriers is in agreement with the distributions of the ambient material and the jet material predicted by a hydrodynamic model apart from a dilute positronic outflow that is accelerated by the electromagnetic field at the jet's head.
\end{abstract}

\maketitle

\section{Introduction}

Accreting black holes emit jets, which are composed of pairs of electrons and positrons and an unknown fraction of ions \cite{Trigo2013}. Their velocity can be moderately relativistic in the case of the microquasars \cite{Falcke96,Mirabel99,Stirling01}. Supermassive black holes in the centers of active galactic nuclei \cite{Madejski16} can accelerate the jet plasma to ultrarelativistic speeds. Some of these jets are collimated, which allows them to cross astronomical distances at a relativistic speed. The fast-moving inner part of the jet remains separated from the surrounding material, which implies that the jet has an internal structure. Their internal structure was discussed for example by \cite{Marti97,Marti03,Leismann05,Bromberg11,Kumar15}. 

A hydrodynamic model was proposed in Ref. \cite{Bromberg11}, which described the structure of a relativistic pair jet that propagated into ambient material with a larger mass density. In the case of microquasars the latter can be the interstellar medium (ISM) \cite{Ferriere01} or stellar wind. Reference \cite{Bromberg11} assumed that the jet is cylindrically symmetric. The jet expelled the ambient material in its way and carved out a path along which the jet material could move freely. Their model considered collimated and uncollimated jets. We focus on the collimated jets, which form if the interaction between the jet and the ambient material is strong. 

Reference \cite{Bromberg11} considered a jet with a planar head that propagated along the axial jet direction and had a normal that is aligned with the cylinder axis. The front of the head was the forward shock between the pristine ambient material and the shocked ambient material. The reverse shock on the rear side of the head slowed down the material of the pair jet that flowed towards the head. The slowed-down material entered the inner cocoon. A contact discontinuity separated the inner cocoon from the outer cocoon, which consisted of the shocked ambient material. Its high thermal pressure led to a lateral expansion of the shocked material on both sides of the contact discontinuity, which then flowed around the jet. A contact discontinuity separated the outer cocoon from the inner cocoon also on the sides of the jet.

Hydrodynamic jet models assume that the collisionality in the plasma of the jet and in its surroundings is large enough to establish and sustain thin shocks and discontinuities on the spatio-temporal scales of interest. A positron with the energy 1 MeV is slowed down to a nonrelativistic speed on a distance of the order of kiloparsecs by its collisional interaction with the interstellar medium if the latter has a number density of the order of one particle per cm$^{3}$ \cite{Jean09}. Collisions may thus not be able to sustain a contact discontinuity between the jet plasma and the ambient plasma on the scale of a microquasar jet, which is considerably smaller than that. 

It is important to determine to what degree the model proposed in Ref. \cite{Bromberg11} is valid also for pair jets, for which the average time between particle collisions is large compared to the growth time of plasma instabilities. Although the hydrodynamic model will still be valid on the global scale of the jet, electromagnetic instabilities and structures will shape the shocks and discontinuities. 

Indeed we know that electromagnetic fields are present in jets. Spectral properties of the synchrotron emissions of electrons and positrons suggest that the jets are permeated by a magnetic field that is coherent on a large scale with superimposed fluctuations \cite{Lyutikov05}. Three sources have been proposed for them.

Firstly, coherent magnetic fields exist at the base of the jet close to the black hole, where they and the radiation are strong enough to generate huge clouds of electrons and positrons \cite{Ruffini04}. Simulations \cite{Moll10} show how this field extracts energy from the accretion disk and accelerates and collimates the jet outflow with this energy. The jet carries the magnetic field with it. 

Secondly, the magnetic field of the interstellar medium (ISM) \cite{Ferriere01} or of the stellar wind of the companion star is compressed as it crosses the jet's external shock, which results in an incoherent downstream magnetic field \cite{Mizuno11}. 

Thirdly, incoherent magnetic fields are generated by nonthermal plasma distributions in the pair plasma of the jet. Separate plasma populations can move through each other during a time that is short compared to the characteristic time between Coulomb collisions of particles. The interpenetration of the jet plasma with the ISM or with plasma clouds within the jet that move at a different speed gives rise to anisotropic particle velocity distributions and to charged particle beams. They relax via the Weibel \cite{Dieckmann18a,Weibel59} or the filamentation instability also known as the beam-Weibel instability \cite{Califano97,Medvedev99,Silva03,Milos06,Lemoine10,Sironi09,Yalinewich10,Lemoine14,Plotnikov18,Vanthieghem18} that drive strong magnetic fields that are coherent on small scales. Particle-in-cell simulations of cylindrical plasma clouds, which consist of cool electrons and positrons, that propagate through an ambient medium show that beam-Weibel instabilities, mushroom instabilities and kinetic Kelvin-Helmholtz instabilities (See Ref. \cite{Nishikawa16} and references cited therein) develop in and close to the jet.

The coherent large-scale magnetic fields and probably also the small-scale magnetic fields, which result from plasma instabilities, play an important role especially on kinetic scales and they need to be studied further. 

We examine here how a cloud of electrons and positrons interacts with an ambient plasma, which consists of electrons and protons. We select values for the cloud's density, temperature and mean speed for which the ensuing instabilities accelerate protons \cite{Dieckmann18b} and generate strong magnetic fields \cite{Dieckmann18c}. In these works, the ambient plasma and the pair cloud were uniform in the direction orthogonal to the expansion direction. Here we consider a pair cloud with a density that decreases quadratically with the distance from its central axis until it reaches zero. The cloud is truncated at its front. 

We show with a simulation using the relativistic and electromagnetic particle-in-cell (PIC) code EPOCH  \cite{Arber15} how the electrons and protons in the ambient plasma compel a part of the pair cloud to form a magnetized jet that expels the protons in its path.  The magnetic field, which is driven by a filamentation instability, acts as a piston that piles up the protons in the lateral direction; an outer cocoon forms, which is separated by an electrostatic shock from the pristine ambient plasma. The shock propagates at about 3\% of the light speed $c$. A flat head of the jet deflects the ambient protons around it into the outer cocoon. The lateral width of the head is comparable to the wavelength of the filamentation instability and it propagates at the speed 0.15$c$.

The magnetic field of the piston is sustained by a filamentation instability, which develops in a small spatial interval around the piston that contains both ambient plasma and particles of the pair cloud. The instability converts the directed flow energy of the pair cloud into magnetic energy and heat and this interval is thus the inner cocoon. Our pair plasma has a thermal momentum spread that is comparable to its drift speed and a shock between the jet plasma and the inner cocoon would be weak. We could not detect one in the simulation.

An electric field is induced around the magnetic piston. It accelerates the positrons that flow out of the jet. It is also responsible for the deflection of ambient protons around the jet's head. It draws the electrons of the ambient plasma into the jet, where they mix with the jet's electrons. The magnetic piston induces an electric field in the outer cocoon, which pulls jet electrons into it.

The magnetic piston, together with the electric fields around it, distributes the carriers of positive charge into a form that is practically identical to that in Ref. \cite{Bromberg11} apart from the dilute outflow of energetic positrons at the jet's head. We thus observe a jet in a collisionless plasma on a microscopic scale that resembles its macroscopic hydrodynamic counterpart in a collisional medium even though the mechanisms, which shape the jet and its internal structure, are quite different.  

Our paper is structured as follows. Section 2 discusses the numerical scheme of our PIC code and the initial conditions of the aforementioned simulation, which we refer to as the main simulation. The filamentation instability between the pair cloud, which moves at a mildly relativistic speed and has a mildly relativistic temperature, and the ambient plasma plays an important role in the formation of the jet in our main simulation. Section 3 shows how the filamentation instability grows and saturates in a reduced geometry and for the initial conditions of the plasma close to the symmetry axis of the pair cloud in the main simulation. Section 3 also discusses how the boundary conditions affect this instability. Section 4 shows how the jet grows out of the interaction between the pair cloud and the ambient plasma in the main simulation. Its structure and its evolution are explained in terms of the filamentation instability. Section 5 summarizes our results.

\section{The code and the initial conditions of the main simulation}

A PIC code represents the electric field $\mathbf{E}$ and the magnetic field $\mathbf{B}$ on a numerical grid and evolves them with discretized forms of Amp\`ere's law and Faraday's law. The PIC code EPOCH we use solves Gauss' law and the magnetic divergence law to round-off precision. Amp\`ere's law requires the plasma current $\mathbf{J}$ in order to update $\mathbf{E}$ in time. 

PIC codes are based on the kinetic equations and they represent the phase space density distribution of each plasma species by an ensemble of computational particles (CPs). Each particle carries with it a charge and a mass and the charge-to-mass ratio must match that of the plasma species it represents. The current contribution of each CP is deposited on the numerical grid and the global sum over all current contributions yields $\mathbf{J}$, which goes into Amp\`ere's law. After the update of the fields, their values are interpolated to the positions of individual CPs and their momentum is updated with a discretized form of the relativistic Lorentz force equation. 

Our main simulation resolves $x$ and $y$ and the jet we observe will propagate along $y$. A spatially uniform ambient plasma at rest fills the simulation box at the time $t=0$. The density and temperature of its electrons with the mass $m_e$ and protons with the mass $m_p = 1836 m_e$ are $n_0$ and $T_0=$ 2 keV. The plasma frequency $\omega_{p}={(n_0 e^2/\epsilon_0 m_e)}^{1/2}$ ($e,\epsilon_0:$ elementary charge and vacuum permittivity) and $c$ define the electron skin depth $\lambda_s=c/\omega_{p}$ that we use to normalize space. 

The temperature $T_0$ exceeds that of the ISM. However, the ISM will be heated up by radiation from an approaching jet. Choosing a high temperature for the ambient plasma yields a large Debye length $\lambda_D = v_{th,e}/\omega_p$ ($v_{th,e}={(k_BT_0/m_e)}^{1/2}, k_B:$ electron thermal speed and Boltzmann constant), which allows us to resolve space with larger grid cells and hence reduce the computational cost. The ambient plasma will be heated to a temperature $\gg T_0$ and we do not expect that a reduction of $T_0$ would affect significantly the jet evolution.

\begin{figure}
\includegraphics[width=\columnwidth]{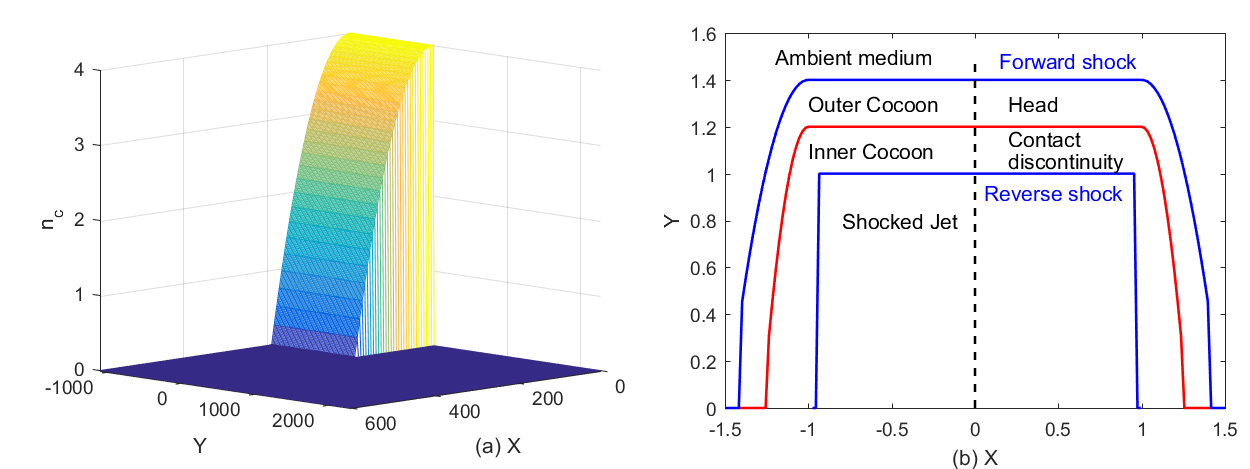}
\caption{Panel(a) shows the initial density distributions of the electrons and positrons of the pair cloud. The boundary condition at $x=0$ is reflecting and forms the symmetry axis of the pair cloud. Panel (b) shows the structure of a relativistic and collimated pair jet in a hydrodynamic model. The dashed line shows the jet's symmetry axis. The pair cloud in (a) and the jet in (b) move along the y-axis.}
\label{fig1}
\end{figure}

We normalize $\mathbf{B}$ to $\omega_p m_e / e$, $\mathbf{E}$ to $\omega_p m_e c/ e$ and densities to $n_0$. The simulation box resolves the interval $0 \le x \le L_x$ with $L_x = 600$ and $-L_y/3 \le y \le 2L_y/3$ with $L_y = 3400$ using a grid with $3500 \times 20000$ grid cells. The electrons and protons are resolved by 28 CPs per cell, respectively. A pair cloud is superimposed on the ambient plasma. The densities of its electrons and positrons are $n_c(x,y)=4-{(x/94)}^2$ for $x \le 188$ and $y\le 0$ and zero otherwise as shown in Fig. \ref{fig1}(a). Each species has the temperature 400 keV and is resolved by $8.4 \times 10^8$ CPs. The mean speed of the electrons and positrons along $y$ is $V_0=0.9c$. The plasma is charge- and current neutral at $t=0$ and we set $\mathbf{E}=0$ and $\mathbf{B}=0$ everywhere. 

Figure \ref{fig1}(b) illustrates the structure of a hydrodynamic collimated jet close to the jet's head in Ref. \cite{Bromberg11}. The jet consists of light material, that moves through material with a larger mass density. 
A collimation shock, which is not depicted in Fig. \ref{fig1}(b), has shocked the pair plasma close to the base of the jet. 

Our simulation setup exploits the symmetry of the jet by slicing it along the dashed line in Fig. \ref{fig1}(b), which results in the distribution of the pair cloud shown in Fig. \ref{fig1}(a). We use reflecting boundary conditions, which cuts the computing time in half and helps us to identify the physical processes at work. Numerical artifacts introduced by the reflecting boundary at $x=0$ are limited to an interval with a width $\lambda_s$ close to the boundary as we show in the next section. The boundary fixes the phase of the filamentation modes that grow closest to the boundary. We stop the main simulation at $t_s=3150$ (unit : $\omega_p^{-1}$) when the fastest particles, which were reflected at the boundary $y=2L_y/3$, return to the region of interest.

\section{Filamentation instability}

We discuss here the growth and saturation of the filamentation instability between the pair cloud and the ambient plasma and how it is affected by the boundary conditions. We compare the results of one PIC simulation with periodic boundary conditions, which we denote as Sim$_p$, with Sim$_r$ that uses reflecting boundary conditions. We argue that the main conclusions of our paper are unaffected by the usage of reflecting boundary conditions, although some differences between both simulations can be spotted on close inspection.

We initialize the plasma in both simulations with the plasma parameters found in the main simulation close to the boundary $x$ = 0 for values $y \le 0$. We resolve only the x-direction, which is the direction that is orthogonal to the expansion direction of the pair cloud. The length 86 of the simulation box is resolved by 500 grid cells. 

All plasma species have a spatially uniform temperature and density to start with. The ambient electrons and protons have the temperature $T_0$ = 2 keV and density $n_0$ and their mean speed is zero everywhere. They are represented by 2000 particles per cell each. The electrons and positrons of the pair cloud have the density $4n_0$, the temperature 200 $T_0$ (400 keV) and they propagate with the speed $V_0=0.9c$ along $y$. Each cloud species is resolved by 3000 particles per cell. All fields are set to zero at $t$ = 0 and we evolve the instability during $0 \le t \le 525$ with $10^4$ steps. This time interval is smaller by a factor 6 than that covered by the main simulation. 

The pair cloud moves with $V_0$ to increasing values of $y$ and a filamentation instability can only displace particles along $x$. Such a displacement results in a current contribution $J_y(x)$ along $y$ of each cloud species. A magnetic field $B_z(x)$ grows if the current contributions of both cloud species do not cancel each other out. An electrostatic field $E_x(x)$ grows if the instability redistributes the plasma species such that space charge is created.

At $t=0$ the CPs were placed at the same positions in Sim$_r$ and Sim$_p$ and the same sequence of random numbers was used to initialize their momenta. We expect at least initially and far from the boundaries an identical evolution of the instability in both simulations. A comparison of the field and particle data provided by both simulations reveals effects introduced by the boundary conditions. 
  
\begin{figure}
\includegraphics[width=0.9\columnwidth]{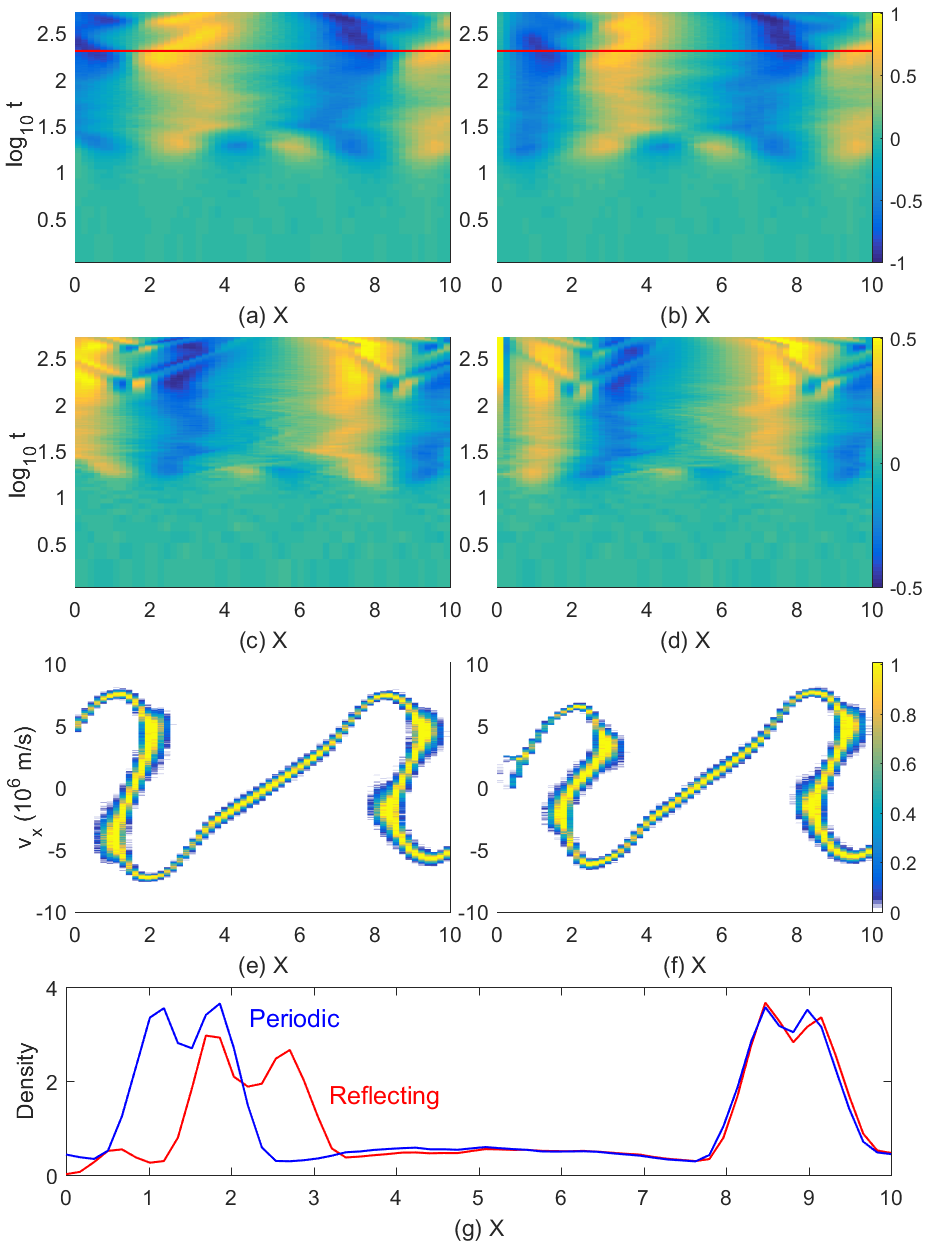}
\caption{The time evolution of $B_z(x,t)$ of Sim$_p$ is shown in panel (a) and that of Sim$_r$ is shown in (b). Panel (c) shows $E_x(x,t)$ of Sim$_p$ as a function of time and (d) that of Sim$_r$. A 10-logarithmic time scale is used in these 4 panels. Panels (e) and (f) show the phase space density distribution $f(x,v_x)$ of the protons in Sim$_p$ and Sim$_r$ at the time $t_c$ = 200 (red horizontal lines in (a, b)), respectively. All color scales are linear and those of the two figures in the same row are identical. Panel (g) shows the proton densities.}
\label{fig2}
\end{figure}

Figure \ref{fig2} compares the time evolution of $B_z(x,t)$ and $E_x(x,t)$ in both simulations as well as the phase space density distribution of the protons in the cut plane $(x,v_x)$ at the time $t_c$ = 200.
Strong fields have developed in Figs. \ref{fig2}(a-d) after the initial exponential growth phase of the instability finished at $t\approx 15$. We can not observe the fields during their initial exponential growth phase since their low amplitudes are not resolved well by a linear color scale. We refer to Ref. \cite{Vanthieghem18} for a discussion of the growth and saturation of a similar instability between counterstreaming pair beams. The protons do not show a strong reaction to the electromagnetic fields at this time (not shown). The field amplitudes remain approximately constant until $t\approx 100$ and they continue to grow afterwards on a slower timescale. 

The phase of the magnetic fields $B_z(x,t)$ in Figs. \ref{fig2}(a, b) differ for the filament closest to $x=0$ but not for the one at larger $x$, which indicates that effects due to the boundary remain localized for these non-propagating waves and for the short time scales we consider. An electric field $E_x(x,t)$ grows in Figs. \ref{fig2}(c, d) that has the same wavelength as the corresponding oscillation of $B_z(x,t)$ and is in antiphase with it. Broad electric field bands surround the spatial intervals $x \approx 2$ and $x\approx 9$, where a drastic change with $x$ of the proton's mean velocity is observed in Figs. \ref{fig2}(e, f). An integration of the phase space density distributions of the protons along $v_x$ yields large density peaks at $x$ = 2 and 9 in Fig. \ref{fig2}(g). These protons have been pushed aside by the electric field. They form phase space vortices \cite{Eliasson06} in the $x, v_x$ plane at later times.

Figure \ref{fig2}(d) reveals the growth of an electric field in the grid cell next to the boundary at $x=0$ until $t\approx t_c$. It remains constant after that. Such a band is not found in Fig. \ref{fig2}(c) and it is thus a consequence of the reflecting boundary. It remains localized and it has no strong effect on the proton distribution in Fig. \ref{fig2}(f). Effects due to the reflecting boundary other than its adjustment of the phase of the filamentation modes are thus weak and limited to a narrow interval.

\begin{figure}
\includegraphics[width=0.9\columnwidth]{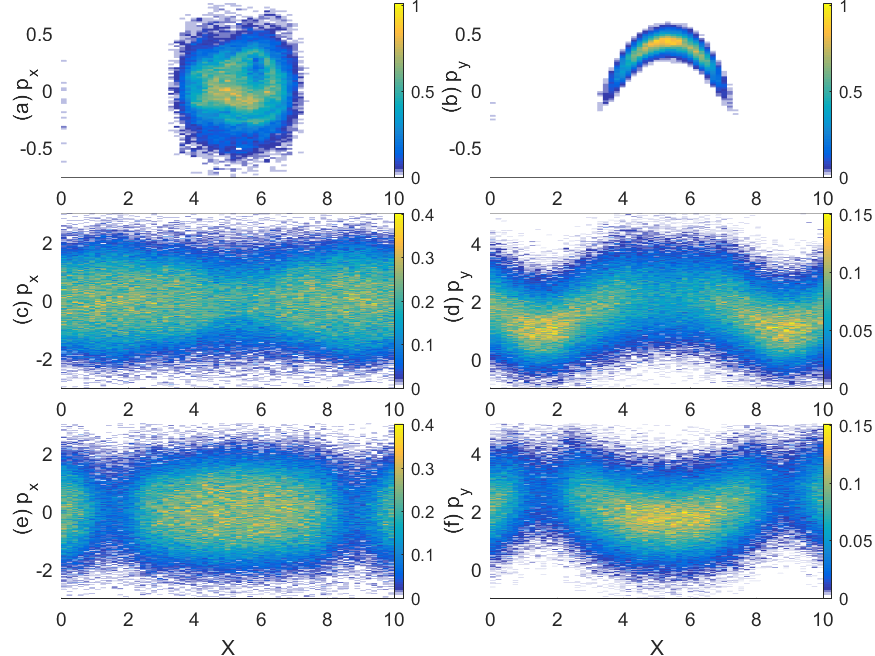}
\caption{Lepton distribution at the time $t_c$ = 200 in the simulation Sim$_p$. The upper row shows the distributions $f(x,p_x)$ (a) and $f(x,p_y)$ (b) of the ambient electrons. The second row shows the distributions $f(x,p_x)$ (c) and $f(x,p_y)$ (d) of the cloud electrons. The bottom row shows the distributions $f(x,p_x)$ (e) and $f(x,p_y)$ (f) of the positrons. All distributions are normalized to the maximum value in the top panel of the respective column and displayed on a linear color range.}
\label{fig3}
\end{figure}

\begin{figure}
\includegraphics[width=0.9\columnwidth]{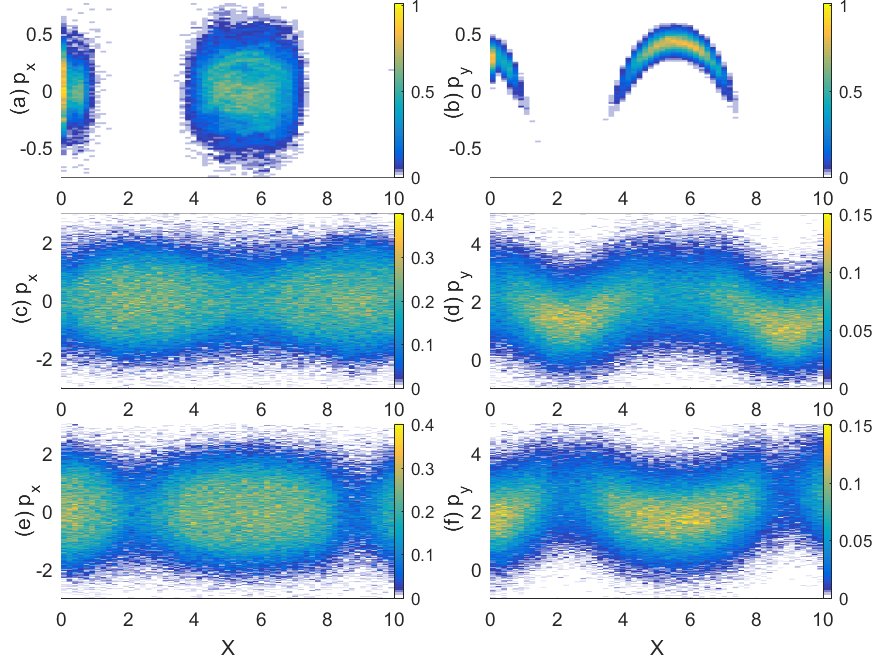}
\caption{Lepton distribution at the time $t_c$ = 200 in the simulation Sim$_r$. The upper row shows the distributions $f(x,p_x)$ (a) and $f(x,p_y)$ (b) of the ambient electrons. The second row shows the distributions $f(x,p_x)$ (c) and $f(x,p_y)$ (d) of the cloud electrons. The bottom row shows the distributions $f(x,p_x)$ (e) and $f(x,p_y)$ (f) of the positrons. All distributions are normalized to the maximum value in the top panel of the respective column and displayed on a linear color range.}
\label{fig4}
\end{figure}

Figures \ref{fig3} and \ref{fig4} show the phase space density distributions of the leptons in Sim$_p$ and Sim$_r$ at $t=t_c$. The projection of the phase space density distribution on the plane spanned by $x$ and by the momentum component $p_x$ is $f(x,p_x)$, while the projection $f(x,p_y)$ involves the momentum component along $y$. The lepton distributions in both simulations are practically identical for $x>2$. They differ close to the boundary, which confines the filament in Sim$_r$ but not that in Sim$_p$. The confinement of the filament closest to $x=0$ in Sim$_r$ does not change qualitatively its lepton distributions. In what follows we only discuss the distributions computed by Sim$_r$.


The ambient electrons in Fig. \ref{fig4}(a) cluster in spatial intervals where the proton density in Fig. \ref{fig2}(g) is low. Their mean momentum along y in Fig. \ref{fig4}(b) reaches a high value close to the boundaries and close to $x=6$ and it decreases to zero at the boundaries of the phase space cloud of the ambient electrons. 

Figures \ref{fig4}(c, d) demonstrate that the cloud electrons accumulate mainly in those spatial intervals with a large proton density. The apparent broadening of the distribution in Fig. \ref{fig4}(c) along $v_x$ at $x\approx 2$ and $x\approx 9$ is at least partially caused by the increased density of the cloud electrons. The distribution of the cloud electrons in Fig. \ref{fig4}(d) reveals that their mean momentum along $y$ is reduced to a non-relativistic value in the intervals where they accumulate. Some of the cloud electrons even reverse their momentum at $x=$ 2 and 9, which implies that the current density of the cloud electrons is close to the Alfv\'en limit (See page 139 in \cite{Miller82}). The positrons in Figs. \ref{fig4}(e, f) have been expelled from the spatial intervals with a high proton density. 

Figures \ref{fig4}(b, d, f) show how electrons and positrons are distributed after the instability saturated in order to drive the current $J_y(x,t)$ that sustains $B_z(x,t)$. We discuss only the distribution close to $x=0$ due to the periodicity of the wave. The positron density has a maximum at $x=0$, while the density of the jet electrons has a minimum at this position. Ambient electrons provide an additional negative current contribution. The mean momentum along $y$ of the positrons increases and their density decreases as we go away from $x=0$. The opposite is true for the electrons. The positron density close to $x=2$ is low and their mean velocity is largest close to this position. The magnetic field is primarily sustained by a spatial separation of both cloud species and its polarity in Fig. \ref{fig2}(b) indicates that the current due to the cloud electrons dominates at $x\approx 2$ while that of the positron cloud is stronger at $x\approx 0$.

The mean momenta in Figs. \ref{fig4}(d, f) remain positive and mildly relativistic. Let us assume that both cloud species drift at the mean speed $\mathbf{v}_D = (0,v_D,0)$ with $v_D>0$ and that this net drift and the magnetic field $\mathbf{B}=(0,0,B_z(x,t))$ drive the electric field $\mathbf{E}_D \approx -\mathbf{v}_D \times \mathbf{B}$ that is observed in Figs. \ref{fig2}(d). The peak values of $B_z(x,t_c)$ and $E_x(x,t_c)$ in Fig. \ref{fig2}(b, d) are 0.8 and 0.5, which gives a resonable value $v_D \approx 0.6c$, and the phase of the drift electric field matches that observed in the simulations. The drift electric field is responsible for the compression of the protons and the filamentation instability is thus not purely magnetic. This has to be expected not only because the center of momentum of the plasma is drifting along y but also because the counter-streaming beams are not symmetric \cite{Fiore06} .

\section{Main simulation}

Here we present the results of the main simulation with the setup we discussed in Section 2 and we show how parts of the pair cloud start to form a collisionless jet that moves along the boundary $x=0$.

\begin{figure*}
\includegraphics[width=0.9\textwidth]{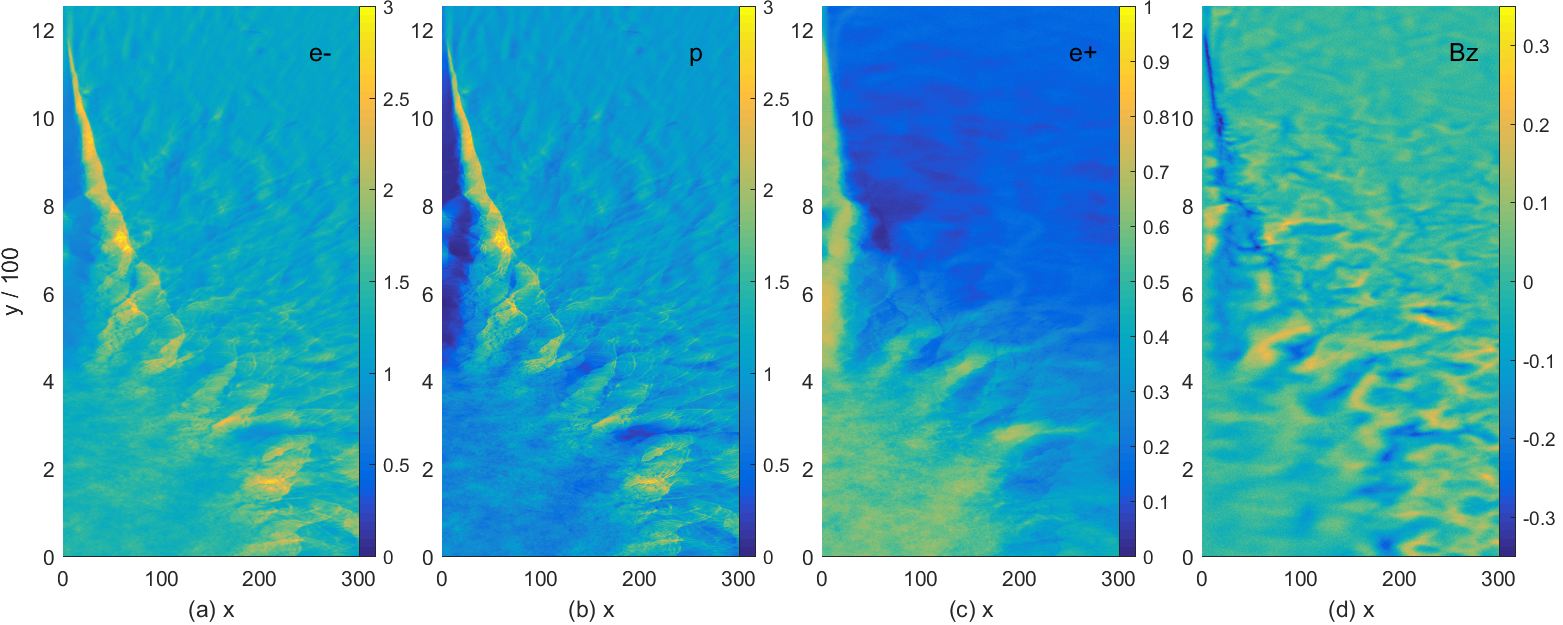}
\caption{The distributions of the plasma densities and of the magnetic $B_z$ component at the time $t_s$: the electron distribution is shown in (a), the proton distribution in (b), the positron distribution in (c) and that of $B_z$ in (d). The color scale in (d) is clamped to values between -0.35 and 0.35. The data has been smoothed with a filter that averages the quantity over $5 \times 5$ cells (Multimedia view).}
\label{fig5}
\end{figure*}

Figure \ref{fig5} shows the density distributions of the electrons, positrons and protons and of the magnetic $B_z$ component in a subinterval of the simulation box at the final simulation time $t_s$. Figures \ref{fig5}(a, b) reveal high-density bands in the distributions of the electrons and protons with the density $\approx 3$ that start at $x\approx 60$ and $y\approx 700$ and end at $x\approx 10$ and $y\approx 1150$. 
No high-density band is present in the positron distribution in Fig. \ref{fig5}(c). The electron and proton densities are close to 1 for values of $x$ that are larger than those of their high-density band and only few positrons are present. 

A magnetic band with the amplitude -0.35 follows in Fig. \ref{fig5}(d) this high-density band and is located at lower values of $x$. It extends to $y < 700$ and maintains its distance $x\approx 30$ from the boundary at $x=0$ for all $450 \le y \le 700$. Oscillatory magnetic fields are observed at $x\approx 200$ and $y<450$. These fields mark the front of the electrons and positrons that expand thermally along $x$. The protons inside the magnetic band have been evacuated from the two intervals $500 \le y \le 700$ and $800 \le y  \le 1150$ and were replaced by positrons with a density between $0.7$ and 1. Two density peaks are observed at $x<20$ for $600 \le y \le 800$ in Fig. \ref{fig5}(b). These structures are composed of electrons and protons. 

The distributions of the plasma species and of the magnetic field in Fig. \ref{fig5} for $500 \le y \le 1200$ resemble a jet, which is propagating to increasing values of $y$ and is centred on the axis $x=0$. In what follows we will refer to the dominant structures in this y-interval with the terms that are defined for the hydrodynamic jet in Fig. \ref{fig1}(b). 

The high-density band in the proton distribution in Fig. \ref{fig5}(b) has no counterpart in the positron distribution and hence it corresponds to piled-up ambient plasma; it is the outer cocoon of the jet. The high-density band observed in the electron distribution in Fig. \ref{fig5}(a) neutralizes its charge. Protons are practically evacuated from the interval to the left of the outer cocoon. This interval is filled with pairs, which constitute the shocked jet material in Fig. \ref{fig1}(b). 

The absence of binary collisions in the PIC simulation implies that contact discontinuities can not be sharp. The outer cocoon is trailed by the magnetic band in Fig. \ref{fig5}(d), which acts like a contact discontinuity. We refer to this discontinuity as the magnetic piston. 

We will now address the processes that resulted in the formation of the jet and interpret them in terms of the filamentation instability discussed in the previous section. 

\begin{figure}
\includegraphics[width=\columnwidth]{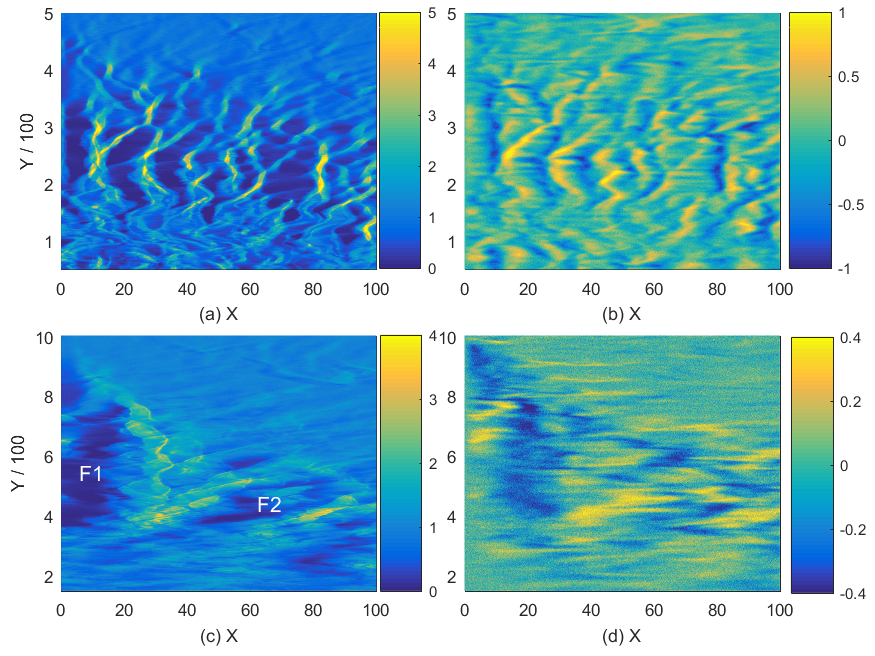}
\caption{Evolution of the plasma density and magnetic field: The proton density distribution at $t=840$ is shown in (a) and that of $B_z$ in (b). The distributions of the proton density and $B_z$ at $t=2000$ are shown in panels (c) and (d) (Multimedia view).}
\label{fig6}
\end{figure}

Figure \ref{fig6} examines the distributions of the proton density and of $B_z$ at the times $t$=840 and 2000. 
Figure \ref{fig6}(a) shows filaments with a thickness $\approx 1$ and separation $\approx 8$, which are followed by magnetic stripes in Fig. \ref{fig6}(b). We focus on the stripe with the value $B_z \approx -1$ at $x\approx 8$ and $230 \le y \le 380$. The magnetic fields in Fig. \ref{fig6}(b) and in Fig. \ref{fig2}(b) have a negative amplitude close to the boundary, which implies that a positive current is flowing along $y$ between them and $x=0$. The proton density in the filament in Fig. \ref{fig6}(a) is low close to the boundary at $x=0$, which matches the density profile in Fig. \ref{fig2}(g), and the peak densities are comparable. 

The proton density close to the boundary in Fig. \ref{fig6}(a) is however lower over a broader spatial interval than that in Fig. \ref{fig2}(g). We understand this difference as follows. The size of the filaments along $x$ is constrained in Sim$_r$ by the small box size and by the presence of equally strong filaments at larger $x$, which is a consequence of the particle distributions that were initially spatially uniform. Filaments can only grow via mergers in the 1D geometry \cite{Vanthieghem18}. The filaments can expand along $x$ in Fig. \ref{fig6} because the pair cloud has a limited extent along $x$. Figures \ref{fig6}(c, d) confirm that indeed the filaments grow in time. 

Figure \ref{fig6}(b) demonstrates that the magnetic field is confined to a narrower interval along $x$ compared to that in Fig. \ref{fig2}(b). A filamentation instability and the currents it drives is constrained to spatial intervals in which positrons and protons coexist since only there a filamentation instability between the ambient plasma and the cloud plasma can develop. The expulsion of protons from extended spatial intervals in Fig. \ref{fig6}(a) suppresses the filamentation instability in these intervals. The filamentation instability could not evacuate all protons in Fig. \ref{fig2}(g) and hence the ambient plasma and the cloud plasma could interact over a broad spatial interval. 

The interval, in which the instability unfolds, has moved to larger $y$ at $t=2000$. Figure \ref{fig6}(c) shows two intervals with a proton density $\sim 0$. The largest depletion is marked with F1. Its right boundary is characterized by a density band with the peak value $\sim 3$ at $x\approx 30$ and a magnetic field $B_z$ with a negative polarity at lower $x$ in Fig. \ref{fig6}(d). The protons have been swept out by the magnetic field structure, which we identify as the magnetic piston in Fig. \ref{fig5}(d). A second similar-sized interval, where protons have been evacuated and replaced by positrons (not shown), is marked with F2 in Fig. \ref{fig6}(c). It demonstrates that the growth of the filament F1 along the boundary is not a numerical artifact. The filament F1 is the dominant one because the ram- and thermal pressures of the pair cloud are largest at low $x$. Indeed Figs. \ref{fig5}(b, c) reveal smaller jets for example at $x\approx 200$ and $y \approx 250$ where protons have been expelled.

\begin{figure*}
\includegraphics[width=0.9\textwidth]{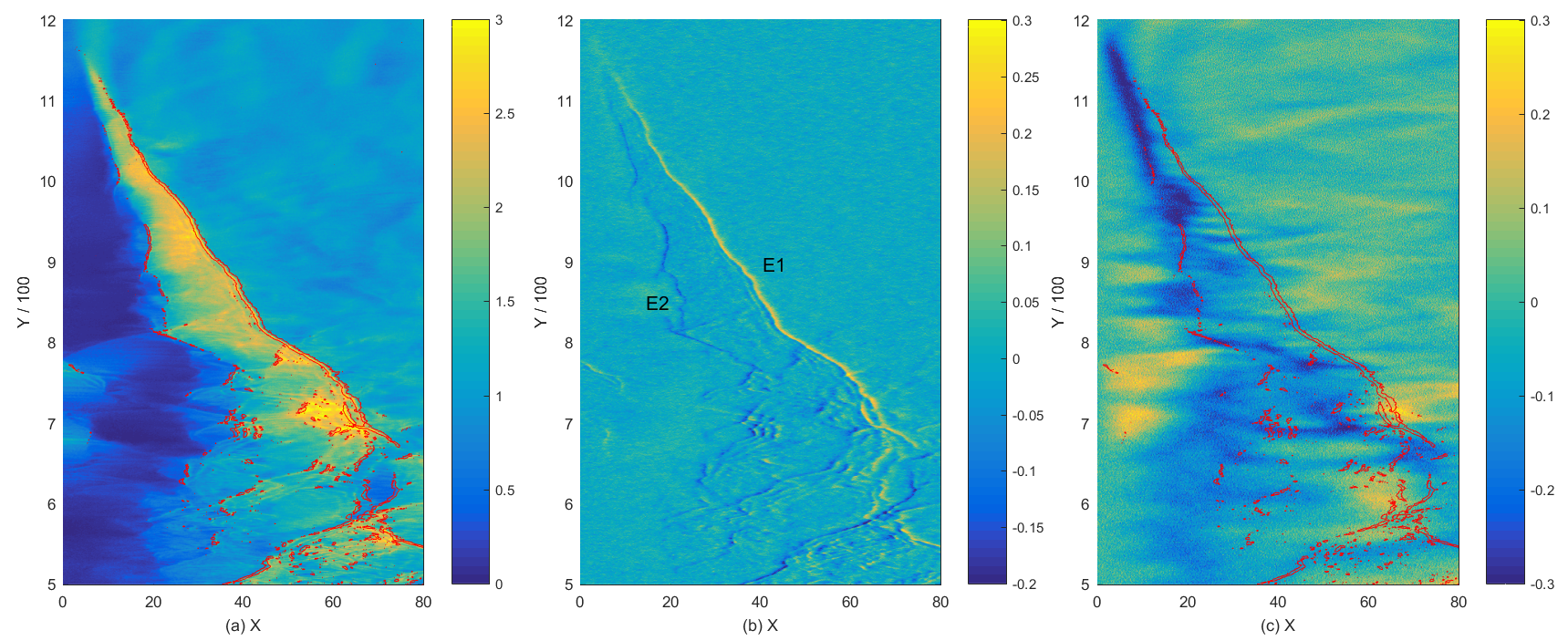}
\caption{The density and field distributions at $t=t_s$: (a) shows the proton density. The electric field component $E_x$ is shown in panel (b) while (c) shows the magnetic $B_z$ component. The contours of the electric field modulus $|E_x|=0.1$ are overplotted in red in (a,c). The band with $E_x = 0.1$ is E1 and that with $E_x=-0.1$ is E2.}
\label{fig7}
\end{figure*}

Figure \ref{fig7} shows the proton density, the magnetic $B_z$ and the electric $E_x$ component in a sub interval of the simulation box close to the largest jet that flows along the boundary $x=0$ at the time $t=t_s$. The other field components are at noise levels. 
The proton density distribution in Fig. \ref{fig7}(a) shows a low-density region at low $x$ with a diagonal front for $800 \le y \le 1150$. This region is bounded to the right by the outer cocoon, which has a thickness that increases approximately linearly along $x$ with decreasing $800 \le y \le 1150$. 

The electric field distribution in Fig. \ref{fig7}(b) shows a narrow electric field band E1 with a positive amplitude and a weaker one E2 with a negative amplitude. Their contours are overplotted in Fig. \ref{fig7}(a) and they match the borders of the outer cocoon. The band E2 follows the minimum of the magnetic field amplitude in the magnetic piston in Fig. \ref{fig7}(c) in the interval $800 \le y \le 1150$.

The magnetic piston extends up to $y\approx 1180$, which marks the jet's head, and it is separated from the reflecting boundary by an interval along $x$ that is comparable to the size of a filament in Fig. \ref{fig2}(b). The magnetic piston does not lead to a sharp outer cocoon at the jet's head. The observed steady increase of the proton density with increasing $y$ across the head is in line with what we expect from a hydrodynamic model. The latter states that the ambient material is piled up and deflected around the head by the forward shock so that it does not enter the jet's interior. The density distributions of the electrons and positrons in the shocked jet material do not reveal jumps and there is no sharp magnetic field band that could mediate a shock between the jet material and the inner cocoon; no well-defined shock has formed that bounds the inner cocoon. 

\begin{figure}
\includegraphics[width=\columnwidth]{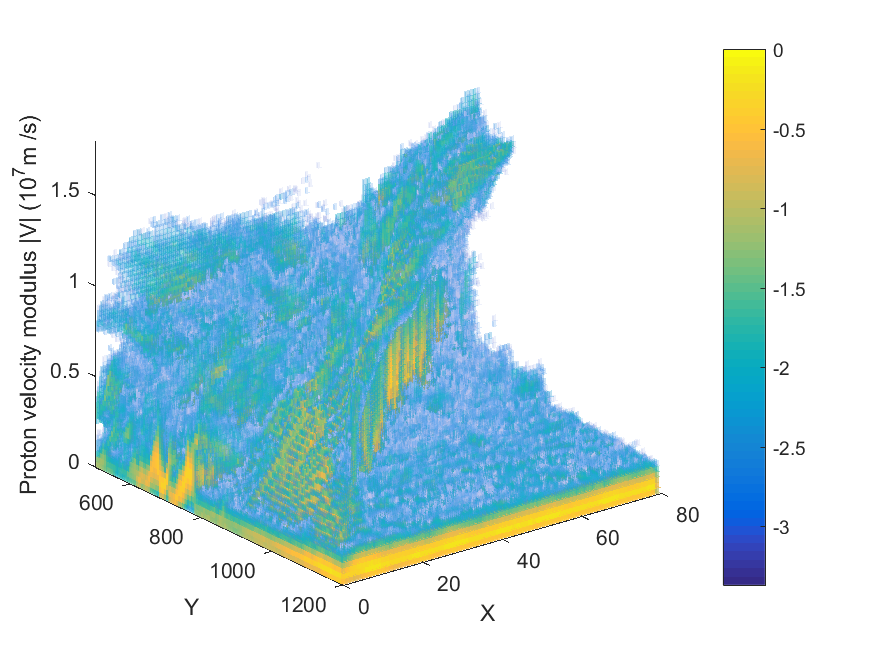}
\caption{The 10-logarithmic phase space density distribution of the protons with $|v|={(v_x^2+v_y^2)}^{1/2}$ at the time $t_s$.}
\label{fig8}
\end{figure}

The phase space density distribution of the protons in Fig. \ref{fig8} reveals the cause of both electric field bands as well as of the proton density spikes in the shocked jet material, which we observed in in Fig. \ref{fig5}(b). The dense spiky structures at $x\approx 0$, $y\approx 700$ and $|v|\le 5\cdot 10^6$ m/s are ion acoustic solitary waves \cite{Jenab17}. These structures form when electrons and positrons stream over protons at rest \cite{Dieckmann18b}. Here they form in the protons that were located behind the magnetic piston when it formed and could therefore not be swept out by it. 

The distribution at large $x,y$ in Fig. \ref{fig8} corresponds to the ambient protons with the temperature $T_0$. The electric field band E1 in Fig. \ref{fig7}(b) sustains an electrostatic shock that starts at $y=1150$ and $x\approx 10$ and extends to increasing $x$ and decreasing $y$. The overturn of the proton distribution into the direction of large $x,y$ at large $|v|={(v_x^2+v_y^2)}^{1/2}$ is typical for a shock-reflected proton beam and a unique feature of collisionless shocks. The shock normal and the direction of the shock-reflected ion beam are almost parallel to the x-axis. We do not observe a shock at the head of the jet, which explains the slow change of the proton density in Fig. \ref{fig7}(a). 

If the shock reflects the protons specularly then the peak value $|v|\approx 1.8 \cdot 10^7$ m/s implies a shock speed $v_s \approx 9\cdot 10^6$ m/s. Figure \ref{fig2}(f) demonstrates that the filamentation instability can indeed accelerate a large number of protons to the speed $v_s$. The ion acoustic speed in the ambient plasma is $c_s = {(k_B(\gamma_e T_0 + \gamma_i T_0)/m_p)}^{1/2}$ with the adiabatic constants $\gamma_e = 5/3$ for electrons and $\gamma_i = 3$ for protons and $c_s \approx 1.7 \cdot 10^6$ m/s. The Mach number of the shock is $\approx$ 5. 

Figure \ref{fig8} does not show a shock that could be associated with the electric field band E2 in Fig. \ref{fig7}(b). That electric field band coincides in space in Fig. \ref{fig7}(a) with a proton density gradient. Its polarity is such that it accelerates protons to lower values of $x$ and $y$, which would erode the density gradient. This is the ambipolar electric field that is driven by thermal diffusion of electrons across a density gradient. The only field structure that is located close to E2 in Fig. \ref{fig7}(b) and strong enough to balance the electric field and, hence, counteract the erosion of the proton density gradient is the magnetic piston with its large amplitude for $B_z$. 

Given the normalization of $\mathbf{B}$ and $B_z\approx -0.35$ we get the electron gyrofrequency $\omega_{ce}=eB_z/m_e\approx \omega_{p}/3$. The proton gyro-frequency $\omega_{ci}=\omega_{ce}m_e/m_i\approx 1.8\cdot 10^{-4}$ and $t_s\omega_{ci}\approx 0.6$. The proton's gyromotion is too slow to be responsible for the plasma's pile-up. 

A large value of $B_z$ also corresponds to a large magnetic pressure. Let us assume that the outer cocoon is at rest and that its electrons and protons have the temperature $T_0$ and density 3. Its thermal pressure is $P_T = 6 n_0k_B T_0$ in SI units. The magnetic pressure is $P_B=\mathbf{B}^2/2\mu_0$ ($\mu_0$: vacuum permeability) in SI units. Only $B_z$ grows in our simulation and we obtain from it the magnetic pressure in SI units as $P_B = \omega_p^2m_e^2 B_z^2 /(2\mu_0e^2)$. We equate the thermal and magnetic pressures and arrive at the condition $B_z^2=12k_BT_0/m_ec^2$ or $|B_z| \approx 0.22$; the magnetic pressure due to $B_z = -0.35$ can balance the thermal pressure of the outer cocoon.

The shock at the front of the outer cocoon forms a straight diagonal line in Fig. \ref{fig7}(a). The shock in \ref{fig7}(a) forms a straight line between $(x,y)=(10,1100)$ and $(40,900)$, which gives the line $y=1100-6.5\cdot (x-10)$. We have previously estimated that the speed of the shock, which bounds the outer cocoon, is about $v_s \approx 9\cdot 10^6$ m/s. The shock propagates perpendicularly to the front of the outer cocoon in Fig. \ref{fig7}(a) and we assume for simplicity that the shock propagates along $x$. 

We have determined the speed of the jet along $y$ by comparing the proton density distribution along $y$ and across the external shock at $x=25$ at the time $0.93t_s$ (not shown) with that at $t_s$. The shock is located at $y\approx 980$ at $x=25$ in Fig. \ref{fig7}(a). We obtained a jet speed $\approx 5v_s$ along $y$. Figure \ref{fig8} shows that the jet's head at $y\approx 1150$ and $x <10$ launches the electrostatic shock, which then propagates along $x$. If the external shock is launched by the head of the jet, which moves about 5 times faster along $y$ than the shock moves along $x$, then we would expect a diagonal front of the outer cocoon with a slope that is comparable to that in Fig. \ref{fig7}(a).

We examine in more detail the mechanism that sustains the magnetic piston and how well it can separate the ambient plasma from the shocked jet plasma. 

\begin{figure*}
\includegraphics[width=\textwidth]{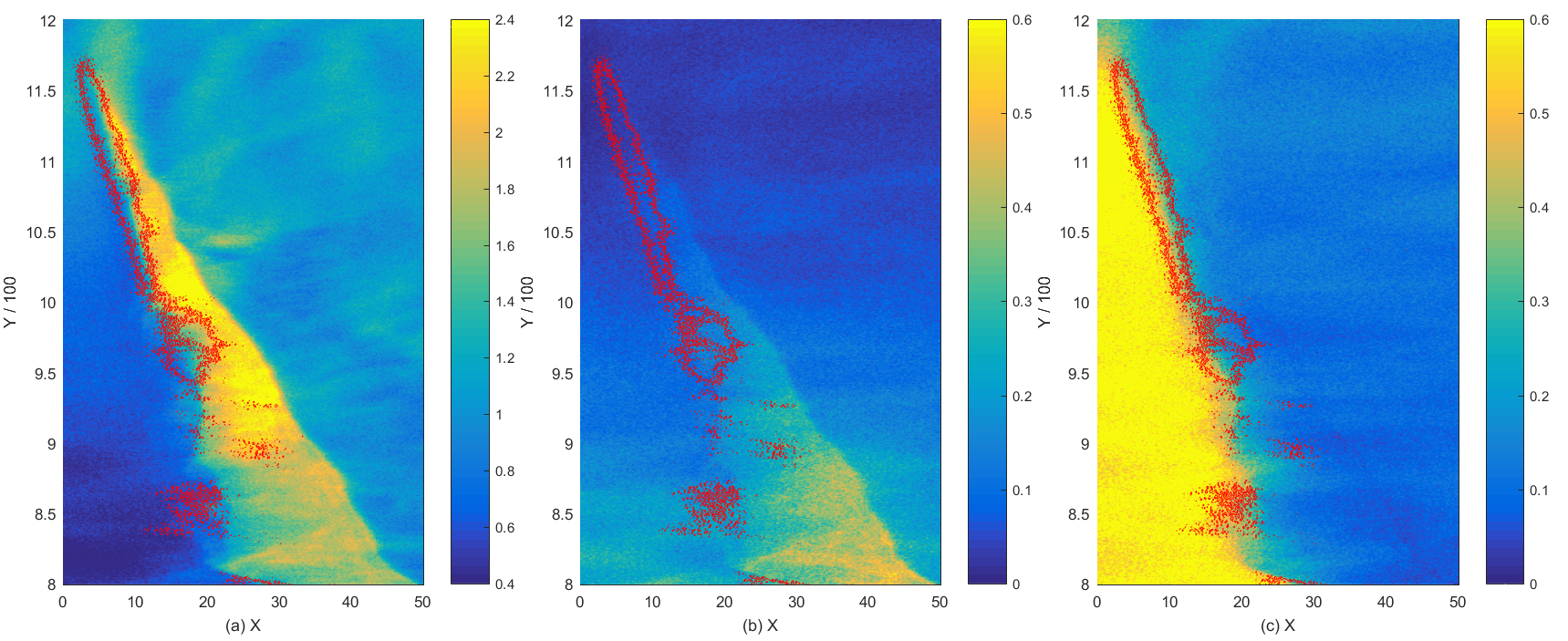}
\caption{The density distributions of the ambient electrons (a), of the electrons of the pair cloud (b) and of the positrons (c) at the time $t=t_{s}$. The color scales are linear. The red curve denotes the contour $|B_z(x,y)|=0.25$.}
\label{fig9}
\end{figure*}

Figure \ref{fig9} compares the density distributions of the ambient electrons and the cloud particles with the contour line $|B_z(x,y)|=0.25$. The contour line marks the position of the magnetic piston.
The outer cocoon is bounded at low $y$ by the magnetic piston. Figure \ref{fig9}(a) demonstrates that most of its electrons are ambient electrons. The ambient electrons also contribute significantly to the shocked jet plasma, which would not be possible in a hydrodynamic model. 

The electrons of the pair cloud in Fig. \ref{fig9}(b) are depleted close to $x=0$ and their density is elevated in the outer cocoon with its increased proton density like in Figs. \ref{fig4}(c, d). Most positrons in Fig. \ref{fig9}(c) are confined to the left of the magnetic piston, which is in agreement with Figs. \ref{fig4}(e, f). We note in this context that only a small fraction of the positrons participated in the jet formation. Most propagated away from the region close to $x=0$ before the jet formed. The confinement of the positrons by the magnetic piston is thus better than that suggested by Fig. \ref{fig9}(c).

We turn to the structure of the jet's head, which is located in the interval $0 \le x \le 4$ and $y\approx 1175$. Its width along $x$ matches the width of the positron cloud close to $x=0$ in Fig. \ref{fig4}(e, f), which suggests that the jet's head is a filamentation mode. The electromagnetic fields of the saturated filamentation mode had a strong effect on the particle momentum distributions in Fig. \ref{fig4}. 

\begin{figure}
\includegraphics[width=\columnwidth]{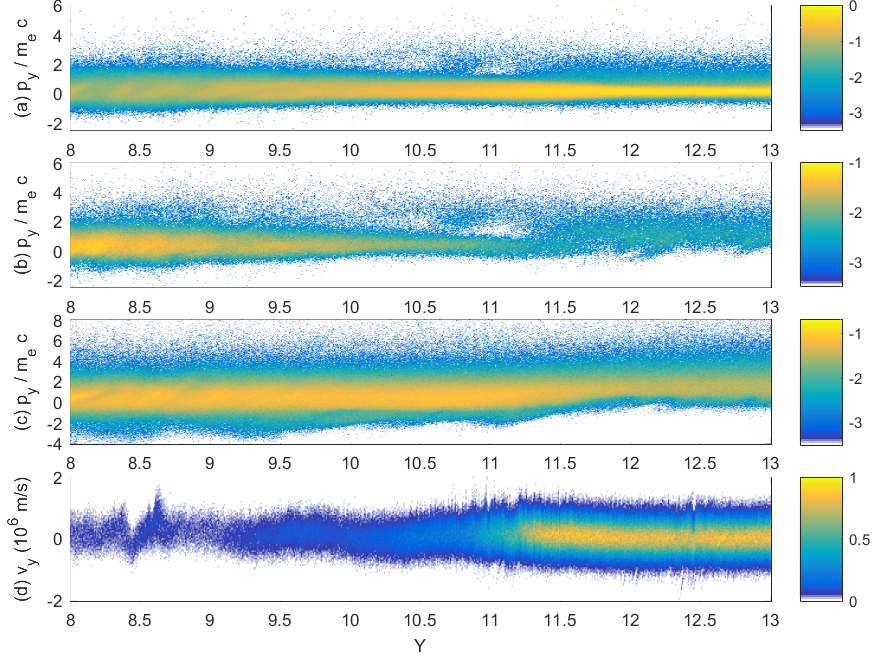}
\caption{The phase space density distributions of the plasma species along the jet propagation direction at the time $t=t_{s}$: The momentum distributions of the ambient electrons (a), pair cloud electrons (b), and positrons (c) are normalized to the maximum value in (a) and displayed on a 10-logarithmic scale. Panel (d) shows the proton velocity distribution on a linear scale. All distributions have been integrated over $0 \le x \le 7.5$.}
\label{fig10}
\end{figure}

Figure \ref{fig10} shows the momentum distributions of the electrons and positrons in the phase space plane defined by $y$ and $p_y$ and the velocity distribution of the protons in the plane spanned by $y$ and $v_y$. We have integrated them across the interval $0\le x\le 7.5$, which includes the magnetic piston.

A comparison of the phase space density distributions of the ambient and cloud electrons along $y$ and $p_y$ in Figs. \ref{fig10}(a, b) shows that both agree apart from the  density change with $y$; the jet electrons are gradually replaced by ambient electrons with increasing $y$. Both species have the same mean momentum and momentum spread along the jet propagation direction. They have thermalized and form one population within the jet. The density of the combined electron distributions is comparable to that of the positrons (See also Fig. \ref{fig9}). 

The positrons in Fig. \ref{fig10}(c) are hotter than the electrons up to $y\approx 1100$ and the fastest ones reach a Lorentz factor that is higher than that of the electrons by 2-3. The mean momentum of the positrons increases in the interval $1100 \le y \le 1200$ where the head of the jet is located. Positrons, which have escaped from the jet via the head, reach a peak momentum $\approx 8 m_ec$ while that of the electrons is $\approx 5 m_e c$. 

Some positrons in the interval $y \le 10^3$ stream back at a speed that yields the Lorentz factor 3. We also find backstreaming electrons but they move at a lower speed. Jet particles, which can keep up with the jet, must move at least with the speed $0.15$c along $y$. Backstreaming particles must have interacted with the jet's electromagnetic fields. Backstreaming particles could only flow along the inner cocoon in a hydrodynamic model. This is not necessarily the case in a collisionless plasma, where no strong constraints exist for the shape of the phase space density distribution along the velocity direction. Figures \ref{fig10}(a-c) evidence backstreaming particles even at $y\approx 800$, where the integration interval $0 \le x \le 7.5$ ends far from the magnetic piston. Here they mix with the material of the shocked jet in Fig. \ref{fig1}(b).

The acceleration of positrons in the interval $1100 \le y \le 1200$ can only be accomplished by an electric field. The protons also react to this electric field. Those at $y \approx 1125$ are faster than those upstream with $y>1250$. The mean velocity of the protons decreases with increasing $y > 1125$ while that of the positrons increases with increasing $1100 \le y \le 1200$. This implies that the electric field structure moves to increasing values of $y$ at a speed that is small compared to that of the positrons. We had previously determined that the head of the jet moves at a speed 0.15 c, which fulfills this criterion. The electric field is thus tied to the jet's head. 

We attribute the electric field $E_y>0$ to a lossy jet front. Amp\`ere's law is $\nabla \times \mathbf{B} = \mu_0 \mathbf{J} + \mu_0 \epsilon_0 \partial \mathbf{E}/ \partial t$. The magnetic term and the current term will balance each other if no current is dissipated. However, current can be dissipated for example by beam instabilities or by the loss of positrons along the normal of the jet's head. An electric field grows in this case which tries to increase the current to the value needed to balance $\nabla \times \mathbf{B}$. If the positive current to the left of the magnetic piston is dissipated away then an electric field $E_y>0$ grows that will accelerate the jet positrons and the ambient protons in the way we observed in Fig. \ref{fig10}. An electric field $E_y>0$ at the jet's head will also accelerate ambient electrons into the jet. Figure \ref{fig9} shows that indeed we find mostly ambient electrons close to the jet's head.

Figure \ref{fig10}(d) shows that, apart from the slight acceleration at the head, the protons maintain their mean speed in the displayed y-interval. We do not observe an electrostatic shock at the front of the jet's head. If there is no shock, then the proton density gradient across the head can only be explained by a proton flow around the head and into the outer cocoon. Such a proton deflection by the jet's head matches the deflection of the ambient material by the head of a hydrodynamic jet. 

The proton deflection can be accomplished in three ways. If the current, which sustains the piston, is dissipative also at its front then the electric field will deflect positive charges around the head and into the outer cocoon. A second contribution comes from the convectional electric field associated with the magnetic field of the piston that moves at 0.15c into the ambient plasma ahead of the jet. A third contribution may also arise from the electric field, which is associated with the magnetic field of the filamentation mode (See. Fig. \ref{fig2}(b, d)).

\section{Summary}

We examined with a PIC simulation the expansion of a pair cloud into an electron-proton plasma. Their large temperature implied that the electrons and positrons of the cloud expanded rapidly, which decreased the density of the cloud. A filamentation instability developed between the ambient plasma and the pair cloud in the interval where the latter was still dense. This instability expelled the protons from large areas, which were then filled with positrons. Magnetic fields grew only in those locations where protons and rapidly streaming jet particles were present, which confined the magnetic field to small spatial intervals. We observed in the simplified one-dimensional study that the filamentation instability drives an electric field. The effect of the electric field is to push protons away from the positron filament. The instability and the magnetic field it drives follows the protons and, hence, the filament grows in size. 

The largest filament grew along the reflecting boundary of our simulation and the magnetic field that swept the protons out became a stable magnetic piston. This filament was the largest one because the density of the cloud was largest close to the boundary and because it was aligned with the flow direction of the pair cloud. It had available the largest pool of directed flow energy, which is converted into magnetic energy by the filamentation instability. 

The filament evolved into a pair jet that was separated magnetically from the expelled and shocked ambient plasma. The front of the jet propagated with the speed 0.15c along the boundary and expanded laterally at a speed that amounted to 0.03c. The growth of the filament was limited by our simulation box size and by the limited cloud size; a decrease of the ram pressure would inevitably lead to a weakening of the filamentation instability and to a collapse of the jet. But it appears that, as long as the pair cloud has enough ram pressure, the filaments can grow to arbitrarily large sizes if the filamentation instability develops between a pair cloud and an electron-proton plasma at least for plasma parameters similar to those we used here. 

We compared the structure of our jet (Fig. \ref{fig7}) with that of a hydrodynamic jet (Fig. \ref{fig1}(b)) and we found several similarities. The magnetic piston acts as a discontinuity between the outer cocoon of the jet and the inner one. Hydrodynamic models attribute this role to a contact discontinuity. Sharp contact discontinuities can only develop in a collisional plasma and hence we could not observe such a discontinuity here. The magnetic piston balanced thermal against magnetic pressure and it was thus similar to a tangential discontinuity. 

Hydrodynamic jets in the model proposed by Ref. \cite{Bromberg11} and the jet in collisionless plasma we observed here have a flat head. We note here that not all jet models predict a flat head \cite{Marti03}. The size of the head was comparable to one wavelength of the waves driven by the filamentation instability. It amounted in our simulation to about 4-5 electron skin depths. If we take into account the fact that our simulation resolved only one half of the head, its width will be of the order of 10 electron skin depths.  It is likely that this width will vary with the ratio between the densities of the pair cloud and the ambient plasma and the temperatures of both. Like for hydrodynamic jets the jet in our simulation deflected the ambient protons around its head and into the outer cocoon. 

The inner cocoon in a hydrodynamic model is defined as the region, where the jet converts kinetic energy into thermal pressure. The thermal pressure pushes the contact discontinuity and the shocked ambient material away from the center of the jet. This was accomplished by our jet in the interval where protons and streaming positrons coexisted and hence we can interpret this region as the inner cocoon. We could not detect a shock between the inner cocoon and the jet plasma. The pair plasma was probably too hot to yield a strong shock and the backstreaming pairs mixed with the pair plasma in the center of the jet instead.

The distributions of the carriers of positive charge resembled those of the light and heavy material in leptonic hydrodynamic jets. We found only one difference: a dilute population of positrons, which was accelerated by the electric field of the jet's head, escaped from the jet into its upstream region. 

Our test simulations demonstrated that the reflecting boundary condition does not lead to artifacts apart from fixing the phase of the filamentation modes. It cut the simulation time in one half and it helped us to determine the various parts of the jet and connect the particle and field distributions to those of the basic filamentation instability. We expect that instabilities may twist the jet in the absence of the rigid spine. Future work should thus address the stability of a jet that does not have a rigid jet spine in the form of a reflecting boundary condition. Future work should also determine if an inner cocoon forms if the pair cloud is colder than the one here.

\textbf{Acknowledgements:} M. E. D. acknowledges financial support by a visiting fellowship from the \'Ecole Nationale Sup\'erieure de Lyon, Universit\'e de Lyon. DF and RW acknowledge support from the French National Program of High Energy (PNHE). GS and MB wish to acknowledge support from EPSRC (grant No: EP/N027175/1). The simulations were performed with the EPOCH code financed by the grant EP/P02212X/1 on resources provided by the French supercomputing facilities GENCI through the grant A0030406960.

\end{document}